\documentclass[letterpaper]{sfchem}
\usepackage{graphicx}

\begin{document}

\title{A database of molecular line data for rotational transitions from selected species of astrophysical interest}

\author{Fredrik L.\ Sch\"oier\inst{1} \and 
            Floris F. S. van der Tak\inst{2} \and
	   Ewine F. van Dishoeck\inst{1} \and
            John H. Black\inst{3}} 
  \institute{Leiden Observatory,  
  P.O.\ Box 9513, 2300 RA Leiden, The Netherlands 
\and
                   Max-Planck-Institut f{\"u}r Radioastronomie, Auf dem H{\"u}gel 69, 53121 Bonn, Germany\and 
                   Onsala Space Observatory, SE-439 92 Onsala, Sweden
} 
\authorrunning{Sch\"oier et al.}
\titlerunning{A database of molecular line data}

\maketitle 

\begin{abstract}
Molecular line data for the rotational transitions of a number of
astrophysically interesting species have been collected and are 
made publically available through the www. These data, including energy
levels, statistical weights, Einstein $A$-coefficients and collisional
rate coefficients, are essential input for non-LTE molecular radiative transfer
programs.  A computer program for performing statistical equilibrium 
calculations is made publically available for use as well. This database should form
an important tool in analyzing observations from current and future infrared and
(sub)millimetre telescopes.
\keywords{Molecular data -- Radiative transfer}

\end{abstract}

\section{Introduction}
A wide variety of molecules has been detected in space to date, ranging from simple molecules like 
CO to more complex organic molecules including ethers and alcohols.
Observations of molecular emission at millimetre and infrared wavelengths, 
supplemented by careful and detailed modelling, are powerful
tools to investigate the physical and chemical conditions of astrophysical objects (e.g., Black 2000). 
For some objects, lines with a large range of critical densities and excitation temperatures are needed, since the densities 
can range from $\sim10^2-10^9$\,cm$^{-3}$ and the temperatures from $\sim 10-1000$\,K
in the interstellar and circumstellar environments probed by current and future instrumentation.

To extract astrophysical parameters, the excitation and radiative transfer of the lines needs to be calculated.
A number of radiative transfer codes have been developed for the interpretation of molecular line
emission [see van Zadelhoff et~al.\ (2002) for a review]. 
The radiative transfer analysis requires molecular data in the form of energy levels, statistical weights and transition frequencies as well as
the spontaneous emission probabilities and collisional rate coefficients. 

\section{Molecular line data}
Literature data on the rotational transitions of 23 different molecules are summarized and extrapolation of collisional rate coefficients to higher energy levels and temperatures are made. The data are made available for the community through the www\footnote{\tt{http://www.strw.leidenuniv.nl/$\sim$moldata}}.
Many of the data files presented here were adopted by Sch\"oier et~al.\ (2002) to model the circumstellar environment of the protostar IRAS 16293--2422.

\subsection{Energy levels and radiative rates}
The energy levels and Einstein $A$-coefficients are obtained from the JPL\footnote{\tt{http://spec.jpl.nasa.gov}} and 
CDMS\footnote{\tt{http://www.cdms.de}} catalogues. 
In some cases, better or more recent laboratory data have been used instead.
Generally, only the ground vibrational state is retained, although in
a few cases, e.g. CO and HCN, vibrationally excited levels are included.
More study of collisional rate coefficients for vibrational
transitions is urgently needed, as there are certainly cases where
vibrationally excited molecules are important. Examples include HC$_3$N
(Wyrowski et~al.\ 1999) and torsionally excited CH$_3$OH.  In the future,
data files including excited vibrational states will be added for more
molecules.

\begin{figure}
\centerline{\includegraphics[height=7.4cm]{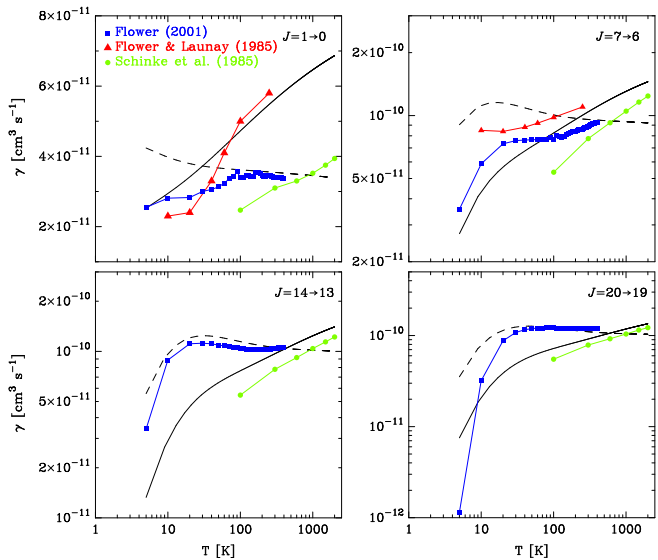}}
  \caption{Calculated collisional de-excitation rate coefficients for CO in collisions with para-H$_2$ for selected transitions (symbols). Black lines indicate fits to the rate coefficients of Flower \& Launay (1985) (solid lines) and Flower (2001) (dashed lines). Also shown are the rate coefficients  presented by Schinke et~al.\ (1985).}
  \label{co}
\end{figure}
\begin{figure}
\centerline{\includegraphics[height=8cm, angle=-90]{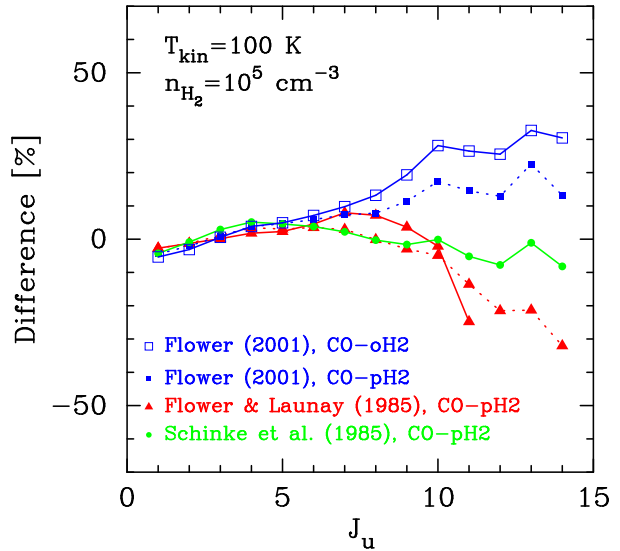}}
  \caption{Predicted CO line intensities using different sets of collisional rates for an isothermal homogeneous sphere with  $T_{\mathrm{kin}}=100$\,K and 
  $n_{\mathrm{H_2}}=10^5$\,cm$^{-3}$, compared to the values obtained using the CO-pH$_2$ rate coefficients from Flower (2001). 
 Points connected with full lines are based on the original sets of rates while the dotted and dashed line styles indicate values obtained from fitting and extrapolation of the original rates}.
  \label{co_mod}
\end{figure}
\begin{figure}
\includegraphics{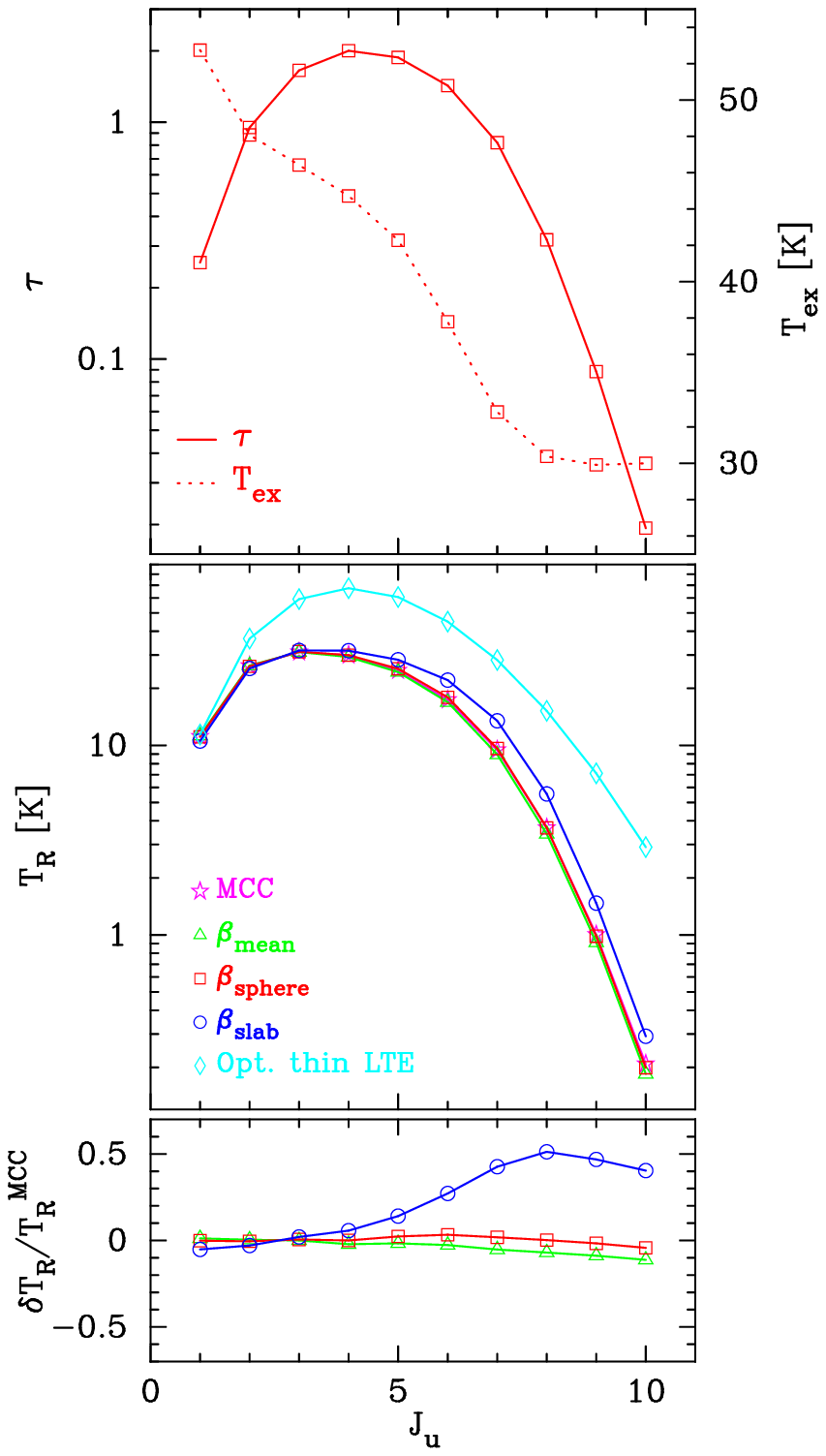}
  \caption{Comparison of the predicted line strengths for the 10 lowest rotational transitions of CO for a homogeneous isothermal sphere, with $n_{\mathrm{H_2}}=10^5$\,cm$^{-3}$ and $T_{\mathrm{kin}}=50$\,K, using different methods. {\em Upper panel} -- The total optical depth through the sphere at line centre $\tau$ and excitation temperature $T_{\mathrm{ex}}$ as a function of the upper rotational level $J$ involved in the transition. 
{\em Middle panel} -- The radiation temperature $T_{\mathrm R}$ obtained for each transition using RADEX with different prescriptions of the escape probability, $\beta$, and compared with the result from the Monte-Carlo code (MCC) of Sch\"oier (2000). 
Also shown are the results for optically thin emission in LTE.
{\em Lower panel} -- $T_{\mathrm R}$ obtained from  RADEX compared with the results from MCC, 
$\delta T_{\mathrm R}=T_{\mathrm R}^{MCC}-T_{\mathrm R}$.}
  \label{compare}
\end{figure}

Hyperfine splitting can be of importance in line transfer
and introduce non-local effects for  lines overlapping in frequency 
(e.g.,  Lindqvist et~al.\ 2000). However, often the splitting between individual hyperfine components is small compared to the line-broadening, so that it can safely  be neglected and treated as a single level for the purpose of excitation analysis. We present data files for some of the relevant molecules, such as HCN and CN, both with and without hyperfine splitting. 

\subsection{Collisional rate coefficients}
The adopted collisional rate coefficients usually pose the largest source of uncertainty of the molecular data input to the radiative transfer analysis. 
Detailed summaries of the theoretical methods and the uncertainties involved in determining 
collisional rate coefficients are given by Green (1975) and Flower (1990).

Published collisional rate coefficients cover only a limited range of temperatures and energy levels, and extrapolations are often necessary (see Fig.~\ref{co}) introducing additional uncertainties. Data files containing the originally calculated set of rates as well as files where an extrapolation has been carried out are available for all molecular species.

In Fig.~\ref{co_mod} the predicted line intensities for the 15 lowest rotational transitions of CO of an isothermal homogeneous sphere are shown using different sets of collisional rates. Deviations up to $\sim 50$\% are found for the higher transitions which are sub-thermally excited. The lower transitions are close to LTE and less sensitive to the adopted set of collisional rate coefficients.

\section{Radiative transfer}
A radiative transfer code, {\em RADEX\footnote{\tt{http://www.strw.leidenuniv.nl/$\sim$moldata/radex.html}}}, is made available for public use as part of the data base.  {\em RADEX} is a one-dimensional radiative transfer code aimed at solving the statistical equilibrium equations using the escape probability
formulation. The code makes no assumption for the geometry or large scale velocity fields and its basic assumption is that of an isothermal and homogeneous medium. 
 {\em RADEX} provides a useful tool in rapidly analysing a large set of observational data providing
constraints on physical conditions, such as density and kinetic temperature.  {\em RADEX}  provides an alternative to the widely used
rotation temperature diagram method (e.g., Blake et al.\ 1987; Goldsmith \& Langer 1999) which relies upon the availability of many transitions of optically thin emission lines and is useful only in roughly constraining the excitation temperature in addition to the column density.

Benchmarking of  the {\em RADEX} code is presented in Fig.~\ref{compare} where various expressions for the escape probability are used. The output is compared both to an optically thin LTE analysis (rotation diagram method) and a full radiative transfer analysis using a sophisticated non-LTE code based on the Monte-Carlo method ({\em MCC;} Sch\"oier 2000). The molecular data used in this example are the collisional rate coefficients for CO collisions with para-H$_2$ from Flower (2001). Levels up to $J=29$ are included. The excitation temperatures of the lines vary from nearly thermalized, for transitions involving low $J$-levels, to sub-thermally excited  for the higher-lying lines. The optical depth in the lines is moderate ($\sim 1-2$) to low. It is seen that the expressions of the escape probability for a sphere and the mean [$(1-e^{-\tau})/\tau$] give almost identical solutions and close to that obtained from the full radiative transfer. However, the slab geometry gives quite different results, in particular for high lying lines. The optically thin approximation, where the gas is assumed to be in 
LTE at 50\,K produces the largest discrepancy and only gives the correct answer for the $J=1\rightarrow 0$ line, where the requirements are fulfilled.
  
\begin{acknowledgements}

FLS and EFvD are grateful  to the Leids Kerkhoven-Bosscha Fond for financial support for their participation in the conference.
This research was supported by the Netherlands Organization for
Scientific Research (NWO) grant 614.041.004, the Netherlands Research School
for Astronomy (NOVA) and a NWO Spinoza grant.

\end{acknowledgements}

\end{document}